\documentclass[sigconf]{acmart}

\usepackage{colortbl}
\usepackage[english]{babel}
\usepackage{graphicx}
\usepackage{amsmath}
\usepackage[super]{nth}
\usepackage{algorithm}
\usepackage{url}
\usepackage{multirow}
\usepackage{tcolorbox}
\usepackage{verbatim}
\usepackage{listings}
\usepackage{caption}
\usepackage{subcaption}
\usepackage{paralist}

\acmConference[ICSE 2024]{46th International Conference on Software Engineering}{April 2024}{Lisbon, Portugal}

\usepackage{arydshln}

\usepackage{chemformula}

\usepackage{arydshln}

\title{Energy Consumption of Automated Program Repair
}

\copyrightyear{2024}
\acmYear{2024}
\setcopyright{acmlicensed}\acmConference[ICSE-Companion '24]{2024 IEEE/ACM 46th International Conference on Software Engineering: Companion Proceedings}{April 14--20, 2024}{Lisbon, Portugal}
\acmBooktitle{2024 IEEE/ACM 46th International Conference on Software Engineering: Companion Proceedings (ICSE-Companion '24), April 14--20, 2024, Lisbon, Portugal}
\acmDOI{10.1145/3639478.3643114}
\acmISBN{979-8-4007-0502-1/24/04}

\author{Matias Martinez}
\email{matias.martinez@upc.edu}
\affiliation{%
  \institution{Universitat Politècnica de Catalunya}
  \city{Barcelona}
  \country{Spain}
}

\author{Silverio Martínez-Fernández}
\email{silverio.martinez@upc.edu}
\affiliation{%
  \institution{Universitat Politècnica de Catalunya}
  \city{Barcelona}
  \country{Spain}
}

\author{Xavier Franch}
\email{xavier.franch@upc.edu}
\affiliation{%
  \institution{Universitat Politècnica de Catalunya}
   \city{Barcelona}
  \country{Spain}
}

\begin{document}

\thispagestyle{plain}
\pagestyle{plain}



\maketitle

\section{Extended Abstract}



\subsection{The problem addressed}

In the last decade, following current societal needs, software sustainability has emerged as research field \cite{calero2015green}.
In this paper, we particularly focus on environmental sustainability, defined as \emph{``how software product development, maintenance, and use affect energy consumption and the consumption of other natural resources. [...] This dimension is also known as Green Software''} \cite{calero2015green}.

The study of environmental sustainability is of paramount importance in the realm of automated software engineering, i.e. the application of computation to software engineering activities with the objective of partially or fully automating these activities, thereby significantly increasing both quality and productivity \cite{Gruenbacher2004ASE-Intro}.

One of these automation-prone activities is bug fixing through automated program repair (APR) tools. This is due to the high economic cost that fixing bugs entails. 
APR tools take as input a buggy program and deliver one or more patches that repair the bugs, when they are found, using test cases as correctness oracle (i.e, to validate candidate patches).
To evaluate their tools, APR researchers usually report the number of bugs that are repaired with 
\begin{inparaenum}[\it a)]
\item a \emph{plausible} patch (that is, a patch valid w.r.t. the correctness oracle), and 
\item  a \emph{correct} patch. 
\end{inparaenum}

However, to our knowledge, \textbf{there is no previous work in the APR area that focuses on the study of the energy consumption of APR tools}.
The APR research community has been oriented towards accuracy, with a strong focus on obtaining state-of-the-art results, as happened in the AI community~\cite{schwartz2020green}.
According to current societal demands and the computational demanding nature of state-of-the-art techniques, we argue that \textbf{APR researchers should also consider energy consumption as a main driver when designing and implementing APR tools}, carefully analyzing whether a minor gain in accuracy justifies high levels of energy consumption.
%

The inclusion of energy as a driver of APR research is important for several reasons.
First, APR tools use expensive correctness oracles, executed each time a candidate patch is synthesized. 
As APR tools synthesize a considerable amount of patches before finding the correct one \cite{Liu2020:ontheefficiency}, the amount of energy up to that point could be considered wasteful.
Second, even with the impressive progress of the field, state-of-the-art APRs do not achieve high accuracy \cite{Jian2023ImpactLLM} and are complementary (as shown, for example, in Fig. 2 from \cite{Ye2022RewardRepair}).
This means that a practitioner aiming to deploy APR would need to consider not just one tool, but several.
Third, connecting APRs with CI/CD (e.g., via a bot~\cite{repairnator}) would require to invoke APRs tools for each failing build caused by test cases with failure or error. 
For example, the study by Moritz et al.~\cite{Moritz2017Test} in Travis CI builds reports a median (resp. mean)  of 2.9\%  (resp. 10.3\%) for builds with failed tests in Java projects.
Triggering APRs for each of these failed builds would require a large amount of repair attempt execution. 

Unfortunately, the impact that APR usage has on the environment is unknown. 
To achieve green software development, we believe it is essential to understand the energy consumption of APR activity.
This work is a first step in that direction.

%

%

\subsection{The approach taken}

The ultimate goal of our research is to:
Analyze \textit{APR tools} with the purpose of \textit{trade-off measurement and analysis} with respect to \textit{energy consumption} from the point of view of \textit{software developers} in the context of \textit{bug repair}.

To achieve this goal, we first measure the energy consumed by APR tools to find plausible and correct patches.
We focus on two families of APRs:
\begin{inparaenum}[\it a)]
    \item \emph{Traditional} tools (i.e., search-, constraint- and template-based), and
    \item \emph{Large-Language Model} (LLM)-based tools.
\end{inparaenum}

Our research is guided by the research question:
\emph{What is the energy consumption of Java APR tools to find the first plausible patch?}
We evaluate on the Defects4J benchmark~\cite{Just2014Defects4J}, the most used bug benchmark, the energy consumption of 10 publicly available traditional repair tools (incl. TBar \cite{Liu2019Tbar}, and SimFix~\cite{Jiang2018SimFix}) and 10 fine-tuned large language models used by Jian et al.\cite{Jian2023ImpactLLM}.
We select these models over other approaches as representative of neural network-based APRs as they achieve state-of-the-art results, beating other neural network (no LLM)-based  approaches such as RewardRepair \cite{Ye2022RewardRepair}.

Once we have the APR tools, we proceed with the experiment as follows.
We measure the energy consumed for each \emph{repair attempt}.
A \emph{repair attempt} is the execution of a repair tool (e.g., TBar) on a particular bug from Defect4J (e.g., Math-70).
We execute the repair attempts on a Linux machine with 8 GPUs connected to an external device called wattmeter which provides the electrical power consumed each 20 milliseconds. 
We compute two timestamps:
\begin{inparaenum}[\it a)]
    \item $ts_0$ of the start of the repair attempt, and 
    \item $ts_p$ of the the first plausible patch is found (after correctness validation). 
\end{inparaenum}
Then, we retrieve from the wattmeter the power samples $Pw= [(ts_0, p_0), (ts_i, p_i), \dots]$ between $t_0$ and $ts_p$. Each power sample $p\_i$ comes with an associated timestamp $ts_i$. 
We compute the energy as follows: 
\begin{equation}
\label{eq:energy}
E_{fp} = \int_{t_0}^{t_p}Pw(t) \approx \frac{(t_n - t_0)}{2} [ p_0 + 2 * p_1 + \dots  +  2* p_{n-1}+ p_n ]
\end{equation}
This give us the \emph{gross} (total) energy consumed to find the patch.
We report the \emph{net} energy, that is, the actual energy spent on the repair, without considering the energy that the machine spends on an idle state (i.e., the energy the node just spends on being turned on).

We report separately the results from the two families of APR (traditional and LLM-based from \cite{Jian2023ImpactLLM})  because, by default, these approaches has different setups, in particular: 
\begin{inparaenum}[\it 1)]
   \item \emph{traditional} uses fault localization, while LLM-based from \cite{Jian2023ImpactLLM} uses perfect fault localization.
   \item LLM-based requires GPUs for LLM inference,
   \item LLM-based APR evaluates exactly 10 candidate patches while traditional stops the evaluation after the timeout or after the first plausible found.
\end{inparaenum}

Finally, since we are interested in the energy of just the first patch ($E_{fp}$), we analyze the correctness of the first plausible patch found in each repair attempt.

\subsection{Preliminary Results}

\begin{table}[]
\centering
\begin{tabular}{| l | r| r|| r|  r| r|  }
\hline
Traditional &&$E_{fp}$&LLM-based&&$E_{fp}$\\
Approach	&	\#P (\#C)	&	Joules	&	Approach	&	\#P (\#C)	&	Joules	\\
\hline											
Avatar	&	44 (15)	&	31694	&	CodeGen350M	&	39 (18)	&	1512	\\
FixMiner	&	21 (9)	&	12995	&	CodeGen2B	&	27 (14)	&	2192	\\
Nopol	&	90 (1)	&	4010	&	CodeGen6B	&	45 (18)	&	3247	\\
\cline{4-6}
Prapr	&	68 (11)	&	3508	&	CodeT5small	&	34 (13)	&	2234	\\
SimFix	&	40 (21)	&	9457	&	CodeT5base	&	42 (13)	&	1680	\\
TBar	&	71 (28)	&	44880	&	CodeT5large	&	23 (10)	&	2189	\\
\cline{4-6}
Dynamoth	&	48 (1)	&	5334	&	InCoder1B	&	44 (13)	&	2253	\\
jGenProg	&	27 (5)	&	5188	&	InCoder6B	&	57 (16)	&	3269	\\
\cline{4-6}
jMutRepair	&	24 (3)	&	3758	&	PLBARTbase	&	34 (14)	&	2433	\\
kPAR	&	36 (9)	&	27663	&	PLBARTlarge	&	41 (12)	&	3006	\\
\hline
\end{tabular}
\caption{Number of plausible patches  (\#P),  where \#C of them are correct, and median values of energy consumed \emph{traditional} tool to find the first plausible patch $E_{fp}$.
}
\label{tab:resultsMetrics}
\end{table}

\paragraph{Traditional APRs}
We observe that TBar, a template-based APR, has the highest median value of energy consumed to find the first plausible patch.
At the same time, it is the tool with the largest number of patches where the first one is correct (28 out of 71 plausible). 
This means that the ability to be the best traditional APR, between the consider ones, is against energy.
We suspect that the reason for both points is its large search space, formed by templates proposed by different previous works.
We also note that other template-based tools such as Avatar, kPAR, FixMiner follow the list of largest electricity consumers to find the first plausible patch.
We highlight the performance of SimFix: the median value is notably lower than those from the mentioned approaches; however, it has the second highest number of correct patches (21 out of 40 plausible patches). 
This shows a good trade-off between energy consumption and repairability.

\paragraph{LLM-based APRs}
We focus on the energy to find the first plausible patch using the setup from \cite{Jian2023ImpactLLM}.
We found that CodeGen350M has, at the same time, the lowest median energy consumed for the first patch and the highest number of correct first patches (18 of 39 patches).
Note that CodeGen350M is one of the ``smallest'' LLM models studied, that is, with the fewest parameters (350 millions).
Other larger models such as CodeGen6B or Incoder6B ($\approx$ 6 billions) arrive to find more plausible patches (but not necessarily more correct patches). 
However, they consume more energy for that because of their hardware requirements (they use 8 GPUs).

\subsection{Conclusion and Future work}

In this paper, we combined two research fields: APR and Green Software.
We presented an initial study reporting the energy needed to find the first patch.
In our research line, we will study other research questions that covers, for example:
\begin{inparaenum}[\it a)]
    \item energy to find the first correct patch (beyond the first plausible)
    \item total energy of the repair process,
    \item dissection of the results according to the type of bugs and project under repair,
    \item relation other with efficiency metrics (e.g, repair time, number of patch candidates).
\end{inparaenum}

\paragraph*{Acknowledges:} This paper has been funded by the “Ramon y Cajal” Fellowship (RYC2021-031523-I) and the GAISSA
Spanish research project (ref. TED2021-130923B-I00;  
MCIN/AEI/10.13039/501100011033).

\bibliography{references.bib}
\bibliographystyle{ACM-Reference-Format}
\end{document}